\documentclass[preprint, secnumarabic, amssymb, nobibnotes, aps, prb]{revtex4-1}

\usepackage{color}
\usepackage{epsfig, graphics, graphicx, subfigure, wrapfig}
\usepackage{verbatim}
\usepackage{dcolumn}
\usepackage{bm}
\usepackage{amsmath, braket}
\usepackage{array}
\usepackage{xr}
\usepackage{filecontents}
\usepackage{tabularx}
\newcolumntype{X}[1]{>{\centering\arraybackslash\hspace{0pt}}p{#1}}
\newcolumntype{M}[1]{ >{\centering\arraybackslash}m{#1}}
\newcommand{\roml}[1]{\lowercase\expandafter{\romannumeral #1\relax}}
\newcommand{\romu}[1]{\uppercase\expandafter{\romannumeral #1\relax}}

\begin{document}

\title{Density functional theory driven phononic thermal conductivity prediction of biphenylene: A comparison with graphene}

\author{Harish P. Veeravenkata}
\author{Ankit Jain}
\email{a\_jain@iitb.ac.in}
\affiliation{Mechanical Engineering Department, IIT Bombay, India}
\date{\today}%

\begin{abstract}
The thermal transport properties of biphenylene network (BPN), a novel $\text{sp}^2$-hybridized two-dimensional allotrope of carbon atoms recently realized in experiments [Fan et al., Science, 372 852-856 (2021)], are studied using the density functional theory-driven solution of the Boltzmann transport equation. The thermal transport in BPN is anisotropic and the obtained thermal conductivities are more than an order of magnitude lower than that in graphene, despite similar $\text{sp}^2$-hybridized planar-structure of both allotropes. The lower thermal conductivity in BPN is found to originate from enhanced anharmonicity which in turn is a result of reduced crystal symmetry of BPN. 

\end{abstract}
\maketitle


\section{Introduction}
Since the successful exfoliation of graphene from bulk graphite using scotch tape \cite{novoselov2004}, there has been a huge surge in research activities in graphene fueling the discovery of its exceptional material properties \cite{novoselov2006, schedin2007, balotin2008, wang2008, geim2010}. Amongst others, graphene has a  high room-temperature carrier mobility of $\sim200,000$ $\text{cm}^2\text{V}^{-1}\text{s}^{-1}$ \cite{balotin2008}, impressive fracture strength of 130 GPa \cite{lee2008a}, optical transparency \cite{nair2008}, ultrabroad optical absorption spectrum \cite{zhang2017}, and extremely high thermal conductivity of $\sim4000$ Wm/mK \cite{balandin2011}. 
Despite the high carrier mobility and other remarkable material properties,  the pristine graphene is a semi-metal which limits its applications in electronics \cite{xia2010}. There is, therefore, a push for the exploration of other carbon-based two-dimensional materials for applications requiring electronic bandgaps such as photo-catalysis,  semiconducting devices, sensors, and thermoelectric energy generation. Many two-dimensional allotropes of carbon, such as graphyne, $\Psi$-graphene, penta-graphene, etc, are now known and they differ from graphene in the nature/arrangement of cyclic rings and associated bonding characteristics \cite{enyashin2011, zhang2015a, li2017}. 

In all of these applications of two-dimensional materials, the material thermal conductivity plays a crucial role in determining the device performance. Since the ultrahigh thermal conductivity in graphene is understood to originate from its planar structure \cite{lindsay2010}, the other planar allotropes of carbon are also expected to result in high thermal conductivity. However, a computational thermal transport study reported thermal conductivity in $\gamma$-graphyne to be 76 W/mK which is an order of magnitude lower than that in graphene \cite{jiang2017}.  The authors suggested that the low thermal conductivity in graphyne is due to the weak bond strength of sp-bonded carbon atoms and lower mass density. In contrast, sp bonds are  generally stronger than $\text{sp}^2$ bonds and the small difference in mass density could not explain the two-orders of magnitude reduction in the thermal conductivity of graphyne compared to that in graphene. 

Yet another computational study on two other computationally-predicted, $\text{sp}^2$-bonded scalar allotropes of carbon, dodecagraphene and tetragraphene, resulted in thermal conductivities to be similar to that in graphyne and 80\% lower than that in graphene \cite{choudhry2019}. The authors attributed the low thermal conductivity of these $\text{sp}^2$-bonded allotropes to superstructure-like modes. These studies suggest that it is not clear a \emph{priori} if other planar allotropes of carbon are going to result in high thermal conductivity similar to that of graphene or reduced thermal conductivity similar to that in graphyne.

\begin{figure}
\begin{center}
\epsfbox{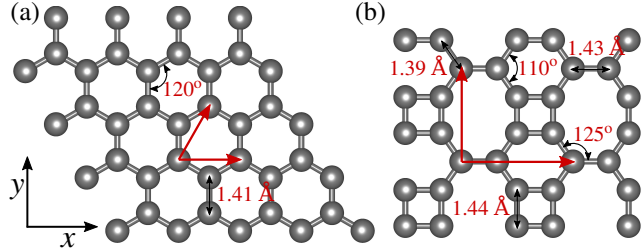}
\end{center}
\caption{The crystal structures of (a) graphene and (b) biphenylene network (BPN). The carbon atoms are arranged in regular hexagons in graphene and  in 4-, 6-, and 8-member rings in BPN. The primitive unitcells of graphene and BPN have 2 and 6 atoms  respectively. All carbon atoms are $\text{sp}^2$-hybridized in both structures, but while all atoms  are equivalent in graphene, they are in two different chemical environments in BPN.}
\label{fig_structure}
\end{figure}

Recently, biphenylene network (BPN), a planar fully $\text{sp}^2$-hybridized allotrope of carbon (Fig.~\ref{fig_structure}), is experimentally synthesized \cite{fan2021}. As opposed to hexagonal rings, the carbon atoms in BPN are arranged in square, hexagonal, and octagonal rings. The studies on the thermal transport properties of BPN, either experimental or computational, are non-existent.

In this work, using ab-initio density functional theory (DFT) driven solution of the Boltzmann Transport Equation (BTE), the thermal transport properties of BPN are investigated and contrasted with graphene. 
The room temperature thermal conductivity in BPN is obtained as  166  and 254 W/mK along the  orthogonal directions. These values are 16-26 times smaller than the thermal conductivity of graphene, despite planar $\text{sp}^2$ hybridized arrangement of carbon atoms in both BPN and graphene. 

\section{Methodology}
The lattice thermal conductivities are obtained using the iterative solution of the BTE and the required harmonic and anharmonic force constants are obtained from the DFT calculations \cite{mcgaughey2019}. As opposed to commonly used relaxation time approximation, the iterative solution of BTE does not treat Normal three-phonon scattering processes as resistive and is found crucial for the correct description of the thermal transport physics in carbon-based materials \cite{lindsay2010, lindsay2014, jain2015}. Whilst the full details regarding the calculation of thermal conductivity from the iterative solution of BTE can be found elsewhere \cite{mcgaughey2019, jain2020}, the thermal conductivity, $k_{ph}$, is obtained as:
\begin{equation}
 \label{eqn_k}
    k_{ph}^{\alpha} = \sum_i c_{ph, i} v_{\alpha}^2 \tau_i,
\end{equation}
where the summation is over all the phonon modes in the Brillouin zone and $c_{ph,i}$, $v_{\alpha}$, and $\tau_i$ represent phonon specific heat, group velocity ($\alpha$-component), and scattering lifetime. 

The harmonic force constants are obtained from density functional perturbation theory calculations (DFPT) with an electronic wavevector grid of $8\times8\times1$ \cite{baroni2001}. The calculations are initially performed on a phonon wavevector grid of $8\times8\times1$ and the obtained phonon properties are later interpolated to $48\times 48 \times 1$ grid for thermal conductivity calculations. The anharmonic force constants are obtained from DFT force-displacement data fitting on 200 thermally populated supercells of  size $6\times6\times1$ (216 atoms) obtained using thermal snapshot technique at a temperature of 300 K \cite{hellman2013}. The DFT forces on supercells are evaluated using Gamma-only sampling of the electronic Brillouin zone. The cubic and quartic interaction cutoffs are set at $5.0$ and $2.0$ $\text{\AA}$ in the force-displacement data-fitting \cite{cubic_cutoff}. The translational invariance constraints are enforced on all extracted force constants. All DFT calculations are performed using planewave energy cutoff of 140 Ry using norm-conserving Von Barth-Car pseudo-potential with local density approximation (LDA) functional as implemented in the  package Quantum Espresso \cite{giannozzi2009} and the phonon thermal conductivity calculations are performed using our in-house code. The layer thickness of $3.4$ $\text{\AA}$ is assumed for the calculation of phonon heat capacity and all reported thermal conductivity includes phonon-boundary scattering with characteristic length of 1 $\text{mm}$.

\section{Results}
The crystal structure of BPN is presented in Fig.~\ref{fig_structure}, along with the structure of graphene. The crystal structure of BPN belongs to the orthorhombic lattice family with a different arrangement of atoms along the $x$- and $y$- directions. The primitive unitcell of BPN has six carbon atoms of which four atoms are bonded to other atoms with bond  lengths of $1.44$, $1.44$, and $1.39$ $\text{\AA}$, while the remaining two atoms are bonded to other atoms with bond lengths of $1.39$, $1.39$, and $1.43$ $\text{\AA}$. The average bond length in BPN is $1.42$ $\text{\AA}$ compared to $1.41$ $\text{\AA}$ in graphene (all carbon atoms/bonds are identical in graphene). The weak bonding of carbon atoms in BPN is also reflected in its reduced atom density ($2.77$ $\text{\AA}^2$/atom compared to $2.58$ $\text{\AA}^2$/atom in graphene) and reduced thermodynamic stability (BPN is 35 mRy/atom higher in energy compared to graphene).

\begin{figure}
\begin{center}
\epsfbox{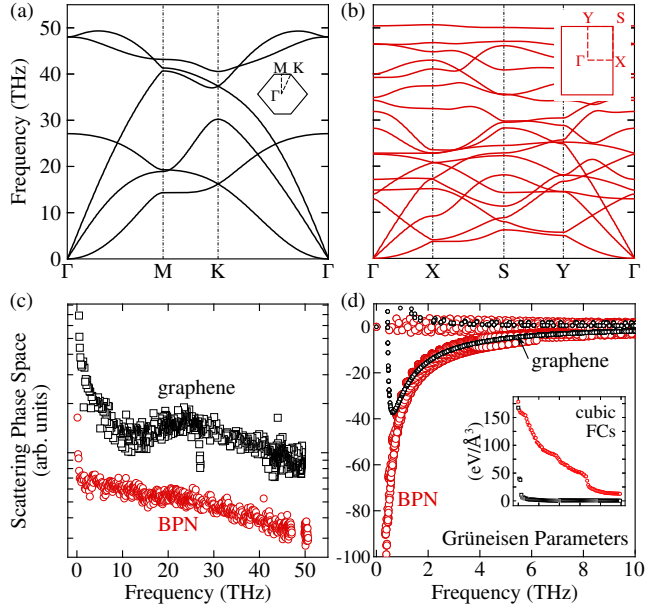}
\end{center}
\caption{The phonon dispersion of (a) graphene and (b) BPN. The graphene and BPN have isotropic and anisotropic dispersions with hexagonal and orthorhombic Brillouin zones respectively. The comparison of (c) three phonon scattering phase-space and (d) Grüneisen parameters for graphene and BPN. The strength of 100 largest cubic force constant matrix elements is plotted in the inset to (d).}
\label{fig_harmonic}
\end{figure}

The phonon dispersion of BPN is presented alongside  that of graphene in Figs.~\ref{fig_harmonic}(a) and \ref{fig_harmonic}(b). Overall, the range of vibration frequencies is similar in BPN and graphene. In contrast to graphene, however, BPN has anisotropic dispersion due to its orthorhombic structure. Further, consistent with weak interatomic bonding and higher energy of carbon atoms in BPN, the group velocities of all three acoustic phonon branches are severely reduced in BPN compared to that in graphene (18500 and 10670 m/s compared to 20650 and 12340 m/s for longitudinal and transverse acoustic phonons in the long-wavelength limit in the $\Gamma-\text{X}$ and $\Gamma-\text{M}$ directions in BPN and graphene). In general, a reduction in group velocity results in a decrease in the thermal conductivity of material (according to Eqn.~\ref{eqn_k}). However, in the case of graphene, the majority of the thermal transport is by flexural phonons having quadratic dispersion and small group velocities. These flexural phonons contribute around 80\% to thermal conductivity due to their reduced phonon scattering rates \cite{lindsay2010}. The possibility and strength of phonon-phonon scattering depend on  scattering selection rules and anharmonicity, characterizable using three-phonon scattering phase space and Grüneisen parameters \cite{jain2014, jain2015b}.

The three-phonon scattering phase space of BPN is compared with graphene in Fig.~\ref{fig_harmonic}(c). The scattering phase measures the fraction of total  three-phonon processes that are able to satisfy crystal momentum and energy conservation selection rules. It is important to note here that since scattering phase-space is normalized by the total number of potential scattering processes [$(3N_a)^2$, where $N_a$ is the number of atoms in the unitcell, i.e., 324 and 36 for BPN and graphene], care must be taken while comparing the rates for different materials. In the case of BPN and graphene, the normalization factor ratio is 9 and as such numbers reported  in Fig.~\ref{fig_harmonic}(c) should be scaled by 9 for BPN to compare them with graphene. After scaling, the scattering phase-space of BPN  is similar to that in graphene for all modes; thus indicating a similar number of phonon scattering events in two materials, though the phonon dispersion of BPN is much more complex compared to that of graphene. 

\begin{figure*}
\begin{center}
\epsfbox{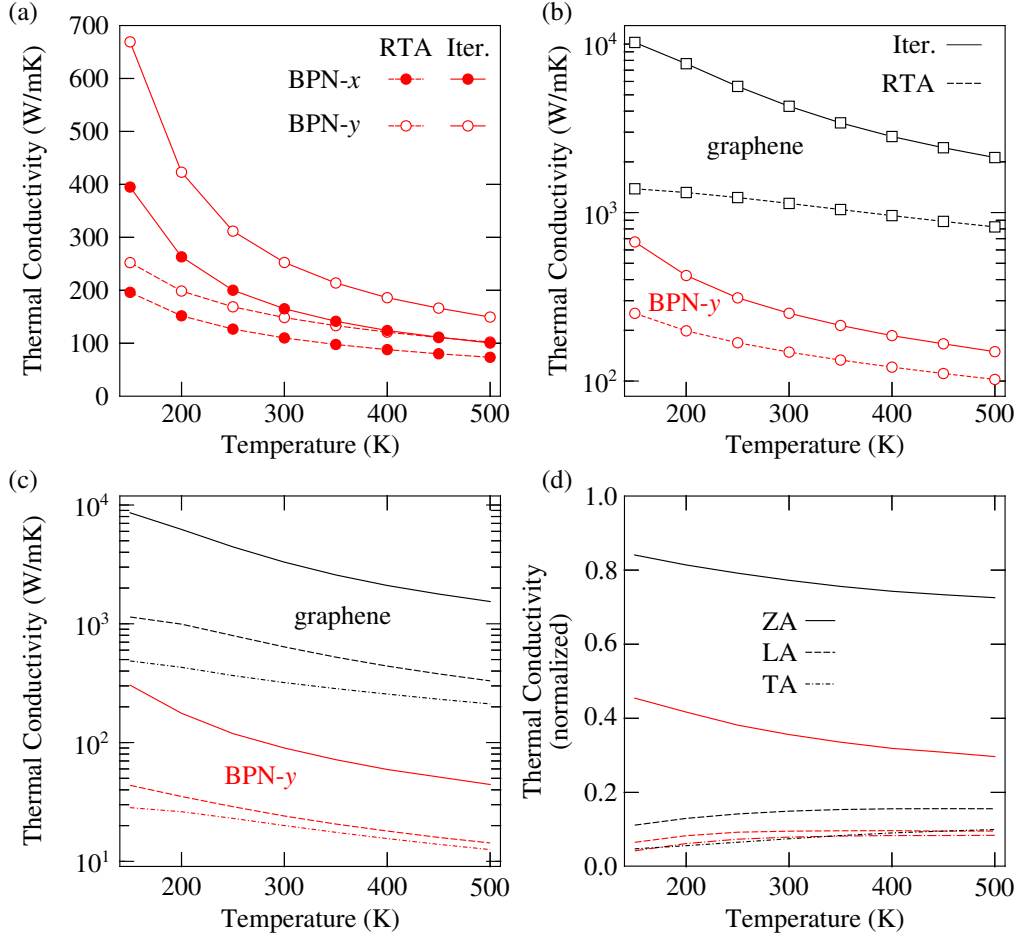}
\end{center}
\caption{The temperature-dependent thermal conductivities as obtained from the relaxation time approximation and the iterative solution of the BTE for (a) BPN in the $x$- and $y$- directions and  (b) BPN ($y$-direction) and graphene. The temperature-dependent contribution of flexural (ZA), transverse (TA), and longitudinal (LA) acoustic modes towards the thermal conductivity on (c) absolute scale and (d) normalized scale. The values in (d) are normalized by the respective total conductivities of BPN ($y$-direction) and graphene. }
\label{fig_k}
\end{figure*}

Next, the anharmonicity of phonon modes of BPN is characterized using the Grüneisen parameters and is compared with graphene in Fig.~\ref{fig_harmonic}(d). As can be seen from Fig.~\ref{fig_harmonic}(c),  modes with frequencies larger than 2 THz have similar Grüneisen parameters in BPN and graphene. However, modes with sub-THz frequencies are significantly more anharmonic in BPN compared to that in graphene. 
To understand the origin of this high anharmonicity of sub-THz modes in BPN, we compared cubic force constants of BPN with graphene in the inset of Fig.~\ref{fig_harmonic}(d). 

Firstly, we find that due to mirror symmetry, all force constants with an odd number of out-of-plane displacements are zero in BPN, similar to that in graphene. It is worthwhile to note here that this zeroing of cubic force constant elements due to mirror symmetry is argued as the reason for the reduced scattering of flexural phonon modes in graphene \cite{lindsay2010}. Next, we find that due to the reduced in-plane symmetry of BPN, the non-zero cubic force constant elements are significantly higher in BPN than in graphene. For instance, for the self-force constant matrix, the number of non-zero elements is 14 (out of a total of 27) in BPN compared to only 4 in graphene. Furthermore, the magnitudes of non-zero force constants are much higher in BPN than in graphene. For nearest  neighbor interaction in graphene, only 3 elements are non-zero and all three of these non-zero elements have values smaller than 40 eV/$\text{\AA}^2$. In contrast, for BPN, 22 nearest neighbor cubic matrix elements have values larger than 10 eV/$\text{\AA}^3$ with two values even larger than 100 eV/$\text{\AA}^3$.

Moving ahead, we next investigate the thermal conductivity of BPN and plot the results as a function of temperature in Fig.~\ref{fig_k}. Due to its asymmetric structure, the thermal transport is anisotropic and we report thermal conductivities in both x- and y- directions in Fig.~\ref{fig_k}(a). The room temperature thermal conductivities are 166 and 254 W/mK and thermal conductivity decrease with increasing temperature due to enhanced Umklapp phonon-phonon scattering at high temperatures. The anisotropy in thermal conductivity is $1.7$ at 150 K and decrease to $1.5$ at 500 K. The lower thermal conductivity in the $x$-direction is understandable from loosely packed atoms with bond-lengths of $1.43$, $1.39$, $1.44$, and $1.39$ $\text{\AA}$ for consecutive bonds compared to stronger bonds with bond-lengths of $1.39$, $1.44$, and $1.39$ $\text{\AA}$ in the $y$-direction.

The thermal conductivity of BPN is compared with graphene in Fig.~\ref{fig_k}(b). Similar to graphene, the phonon-phonon scattering in BPN is dominated by momentum-conserving Normal three-phonon scattering processes. As a result, the correct treatment of phonon-scattering requires an iterative solution of the BTE. The thermal conductivity obtained from iterative solution is a factor of three higher than that from relaxation time approximation at a temperature of 150 K for BPN compared to a corresponding increase by  a factor of seven in graphene. This suggests that even though the scaled three-phonon scattering phase space (and hence the number of scattering processes experienced by phonons) of BPN is similar to that in graphene [see Fig.~\ref{fig_harmonic}(c)], a much larger fraction of these processes are non-resistive in graphene than that in BPN. 

The  thermal conductivity of BPN is a factor of 14-29 lower than that of graphene in the considered temperature range. As discussed earlier, this low thermal conductivity of BPN originates from the strong anharmonicity of bonds in BPN compared to that in graphene. Due to this strong anharmonicity, while the contribution of all phonon branches is lower in BPN compared to that in graphene, the reduction is maximum for flexural acoustic phonons which contribute 3300 W/mK to thermal conductivity in graphene and only 90 W/mK in BPN at a temperature of 300 K [see Fig.~\ref{fig_k}(c)]. The relative contribution of flexural acoustic phonons  to the total thermal conductivity is reduced from 77\% in graphene to 36\% in BPN at room temperature [Fig.~\ref{fig_k}(d)]. For transverse and longitudinal acoustic phonons, the combined relative contribution reduces from 22\% in graphene to $17\%$ in BPN.

\begin{figure}
\begin{center}
\epsfbox{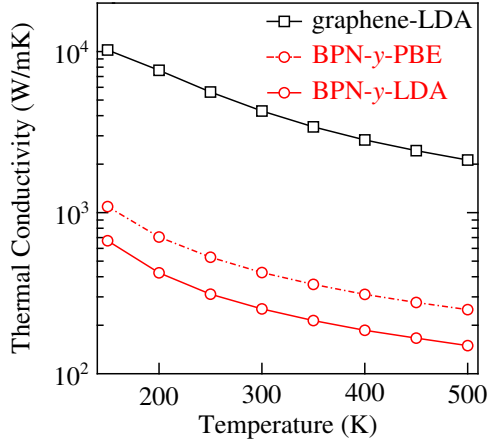}
\end{center}
\caption{The role of exchange-correlation functional on the predicted thermal conductivity (from iterative solution) of BPN ($y$-direction). }
\label{fig_functional}
\end{figure}

Due to the presence of just a monolayer of atoms, the transport properties of two-dimensional materials are more sensitive to lattice strain than the bulk three-dimensional materials. As such, the thermal conductivity of BPN obtained using LDA functional could have a large error due to under-prediction of true lattice constant \cite{jain2015b}. To check for this, we also performed calculations using Perdew-Burke-Ernzerhof (PBE) \cite{perdew1996} exchange-correlation functional. For these calculations, we employed a planewave energy cutoff of 80 Ry, and phonons are initially obtained on a wavevector grid of $12\times12\times1$ using an electronic wavevector grid of size $12\times12\times1$ from DFPT calculations. Other computational parameters are kept the same as those discussed earlier for LDA functional. The relaxed lattice constant from PBE functional is $4.47$  ($3.37$) $\text{\AA}$ compared to $4.51$ ($3.76$) $\text{\AA}$ from LDA functional in the $x$ ($y$) direction. The results obtained with PBE functional are reported in Fig.~\ref{fig_functional}. As can be seen from Fig.~\ref{fig_functional}(b), while there is a 40\% difference in the predicted thermal conductivities from LDA vs PBE functional, the values are still an order of magnitude times lower than that of graphene at all temperatures. 

\section{Summary}
In summary, we used density functional theory calculations to investigate the thermal transport properties of biphenylene network (BPN), which is a planar $\text{sp}^2$ hybridized network of carbon atoms, recently realized in experiments. We find that the thermal transport is  anisotropic in BPN and the predicted thermal conductivities are an order of magnitude lower than that in graphene. The origin of this reduced thermal conductivity is identified as enhanced bond anharmonicity originating from reduced crystal symmetries in BPN. The effect of anharmonicity is maximally felt by flexural acoustic modes whose contribution reduces from 77\% in graphene to 36\% in BPN at room temperature.


The authors acknowledge the financial support from IRCC-IIT Bombay. All the calculations are carried out on SpaceTime-II supercomputing facility of IIT Bombay.

The raw/processed data required to reproduce these findings is available on a reasonable request via email.

%


\end{document}